# Globular Clusters in Mergers

Uta Fritze - v. Alvensleben[1,2], Andreas Burkert[3]

[1] Universitätssternwarte, Geismarlandstr. 11, D - 37083 Göttingen, Germany
[2] Landessternwarte, Königstuhl, D - 69117 Heidelberg, Germany
[3] MPIfA Karl-Schwarzschild-Str. 1, D - 85748 Garching b. München, Germany

**Abstract.** From spectrophotometric and chemical evolutionary modelling of interacting and merging galaxies we predict abundances of burst stars and of globular clusters (GCs) that may be formed in strong bursts. Observations of young GCs in merger remnants confirm the predicted metallicities. We investigate their formation as a result of a strong star burst with a high star formation efficiency over a large volume of the merger remnant.

The evolution of GCs of various metallicities in terms of broad band colours and stellar metal indices are modeled. Comparing our results to the V-I, Mgb and Fe5270 observations of GCs in NGC 7252 largely removes the age – metallicity degeneracy and allows for accurate age dating as well as for estimating the effects of self – enrichment.

**Key words:** galaxies:interactions, evolution, abundances - stars: formation

## 1. Introduction

The difference in the specific globular cluster (**GC**) frequency between spirals and ellipticals has been evoked as an argument against a spiral – spiral merger origin of elliptical galaxies (van den Bergh 1990). Usually, the specific GC frequency $S_{GC}$ in a galaxy is defined as the number of GCs per unit V – luminosity: $S_{GC} := \frac{N_{GC}}{L_V}$ (Harris & van den Bergh 1981). Spiral galaxies (Sp) have $S_{GC}$ in the range $1 - 2$, while ellipticals typically have $S_{GC}$ values a factor of several higher, and up to $S_{GC} \sim 16$ for cluster cDs (van den Bergh 1990, Harris 1991). Recently, Zepf & Ashman (1993) normalised $N_{GC}$ by the stellar mass $M_*$ of the galaxy in units of $10^9 \, M_\odot$ instead of the V – luminosity to be able to account for different $M/L_V$ values for different galaxy types. From observations of samples of both ellipticals and spirals they find that $T_{GC} := \frac{N_{GC}}{M_*/10^9 \, M_\odot}$ is again larger in ellipticals than in spirals with $\langle T_{GC} \rangle_E \sim 2 \cdot \langle T_{GC} \rangle_{Sp}$. If ellipticals result from one major spiral – spiral merger, they conclude that the number of GCs which formed in the merger-induced starburst is comparable to the number of GCs originally present in the two progenitor spirals.

Different GC formation scenarios have been proposed in the literature (e.g. Fall & Rees 1985; Freeman 1990; Murray & Lin

*Send offprint requests to*: Uta Fritze - v. Alvensleben

1992; Brown, Burkert & Truran 1991, 1995; Ashman & Zepf 1992; Kumai et al. 1993; Harris & Pudritz 1994). In all models the globular clusters form from cold gas clouds which condense into a stellar system. The self-enrichment model of Brown et al. (1991,1995) assumes that supernova explosions inside a molecular cloud trigger the formation of a second generation of stars within the same cloud. It is shown that bound GCs would form in this way, if the star formation efficiency (**SFE**), defined as the mass of second generation stars $M_*$ produced from a given gas mass $G$, $\eta := \frac{M_*}{G}$ is unusually high: $\eta \gtrsim 0.1$. Otherwise, the star cluster would be dispersed in the tidal field of the galaxy.

Detailed comparison of spectrophotometric and chemical evolutionary models (Fritze - v. Alvensleben & Gerhard 1994a, hereafter **Pap. I**) with observational data (colours, spectrum, luminosity, mass, gas content) and results from dynamical modelling (Borne & Richstone 1991) for the merger remnant NGC 7252 have led Fritze - v. Alvensleben & Gerhard (1994b) to conclude that – on a global scale – the SFE in this system during the strong burst $\sim 1.3$ Gyr ago should have been very high: $\eta \gtrsim 0.2$ and, more probably, even $\eta \sim 0.45$. On the basis of this high SFE and Brown *et al.*'s results, the formation of a secondary population of GCs ought to be expected.

Young GCs in merger remnants have indeed been found, first in NGC 3597 by Lutz (1991), then by Holtzman et al. (1992) in NGC 1275. In this central galaxy of the Perseus cluster, GC formation has been discussed in the context of a long suspected cooling flow (Richer *et al.* 1993). However, recently several pieces of evidence point towards an accretion event of a massive gas rich object causing a starburst roughly 1 Gyr ago (Zepf & Ashman 1993, Ruby *et al.* 1993). The most convincing and clear-cut evidence for GC formation in the course of a Sp – Sp $\longrightarrow$ E merger was presented by Whitmore *et al.* (1993) who reported the HST discovery of some 40 candidate young GCs within 30 '' of the center of NGC 7252. They give positions, $V-$ and $I-$ luminosities, $(V - I)$ colours, and effective radii $R_{eff}$. Schweizer & Seitzer (1993) present spectroscopy of the two brightest members of this sample.

Fritze - v. Alvensleben & Gerhard (1994a) have argued that the metallicity of a secondary population of GCs should at least be as high as that of the ISM in the merging progenitor galaxies, $Z_{GC} \gtrsim Z_{ISM(Sp)}$. They explored the metallicity range expected for mergers of various spiral types over a large range of galaxy ages. Here, we present evolutionary synthesis models for the evolution of GCs with various metallicities in the range $10^{-4} \leq Z \leq 2 \cdot Z_\odot$, including stellar metallicity indicators like Mgb and Fe5270. We compare our results to abundances observed for young GCs in merger remnants and show that once the age – metallicity degeneracy is removed by chosing the appropriate initial metallicity, very accurate age dating becomes possible.

GCs turn out to be a powerful tool to study the SFE and details of the SF process in merger or accretion events. Their metallicity distribution can give decisive clues about the origin of elliptical galaxies. Abundance ratios, whence observed in GC spectra may provide insight about self – enrichment and probably even about the IMF in starbursts.

## 2. GCs as probes for SFE

Gravitationally bound star clusters form most likely if a large mass fraction of a giant molecular cloud turns into stars. Applying the virial theorem and assuming a rapid destruction of the gas cloud after the star formation event, more than 50 % of the total cloud mass must turn into stars in order for the star cluster to remain gravitationally bound after the cloud is dispersed. The more detailed GC-formation model of Brown et al. (1991, 1995) and Burkert et al. (1993) predicts a minimum star formation efficiency (SFE) of 10 % which is still considerably higher than the presently observed mean SFE in our Galaxy. As typical masses of giant molecular clouds in our Galaxy (non-interacting!!!) are $M_{cloud} \approx 10^5 M_\odot - 10^6 M_\odot$, the high SFE condition leads to globular cluster masses which are in agreement with the observations.

Elmegreen *et al.* (1993) have shown that much larger molecular cloud complexes can form in interacting systems. As discussed also by Ashman & Zepf (1992) and Harris & Pudritz (1994) such supergiant molecular clouds could contain massive, embedded cores which could be the sites for GC formation. Then only a small fraction of the whole cloud complex might end up in a star cluster and the SFE, averaged over the whole cloud could be small. On the other hand, the local SFE of the star forming core must again be high in order for the cluster to remain gravitationally bound.

The high-mass stars of the newly formed cluster will heat-up and destroy the surrounding cold gas cloud through their UV-radiation, stellar winds and supernova explosions on a timescale of $\sim 10^6$ yr. The formation of a globular cluster therefore has to occur on timescales, shorter than $10^6$ yr, requiring a large local star formation rate of $\sim 10^{-7} M_\odot$ yr$^{-1}$ pc$^{-3}$. Note, that the mean SFR, averaged over the whole system can again be significantly smaller if there does not exist a global mechanism (like e.g. a merger of galaxies) which triggers efficient star formation in different gas clouds at the same time.

In summary, two conditions seem to be required for a globular cluster to form: a high, local star formation efficiency (SFE) which occurs on a short timescale. Both conditions are characteristic for star burst regions which therefore represent ideal places for the formation of globular clusters.

Observations in the Milky Way indicate that molecular clouds have typical lifetimes which are of the order of $10^8$ years. As the internal collapse timescale of these clouds is much smaller ($\sim 10^5$ years) they must be stabilized through stellar energy input, leading to a highly turbulent, internal flow. In addition, magnetic fields can prevent the dissipation of the kinetic turbulent energy on a short timescale and can stabilize the high-density cloud cores against gravitational collapse. Larson (1993) proposed that these cores will only be able to collapse and form stars if their densities exceed a critical density $\rho_{crit} \approx 10^4$ cm$^{-3}$. Globular cluster therefore might form, if a large fraction (more than 10%) of a molecular cloud or a massive core inside a supergiant molecular cloud is compressed to densities $\rho > \rho_{crit}$, simultaneously.

One possible mechanism which could induce such an effect in galaxy-galaxy mergers are direct collisions of molecular clouds. In this case the gas is compressed into a high-density sheet which fragments into stars on a short timescale, forming a young globular cluster. Another possibility is the compression of a molecular cloud due to an increase in the ambient gas pressure as a result of a merger (Jog & Solomon 1992). During the merger the gravitational potential and therefore the dynamics of the cloudy system will be strongly distorted, precipitating high-velocity, head-on cloud-cloud collisions. Collisions of diffuse HI-clouds would increase the ambient HI gas pressure of giant molecular clouds, indirectly triggering their collapse.

Note that in both scenarios cloud-cloud collisions would efficiently dissipate the kinetic energy of the cloudy system. This process could trigger an infall of molecular and HI gas into the inner regions of the merger remnant if the gas does not condense into stars on timescales which are shorter than the dynamical timescale of the galaxy.

Given the assumption that GCs form during merging events we expect to find three types of globular clusters inside merger remnants: two old components from the globular cluster systems of the two progenitors and a young, probably more centrally concentrated component which formed as a result of the merging (Ashman & Zepf 1992). Whereas the old clusters will usually be metal-poor, the new component should consist of preferentially metal-rich globulars which formed from the metal-enriched gas inside the merging galaxies.

## 3. Metallicities of GCs in Sp – Sp Mergers

In **Pap. I** the metallicity range of the ISM in merger progenitor spirals of various types and evolutionary ages was explored. Using simple 1 – zone 1 – phase models, the different spiral types $Sa$, $Sb$, $Sc$, $Sd$ were described by star formation (**SF**) laws with different characteristic timescales $t_* = 2$, 3, 10 and 16 Gyr, respectively. We find that for $Sb$ through $Sd$ galaxies with ages from 9 to 15 Gyr the global ISM metallicities $Z$ range from $1/3 \cdot Z_\odot$ to about $Z_\odot$. Only $Sa$ galaxies at ages $\gtrsim 10$ Gyr reach $Z > Z_\odot$; late mergers of two of these gas poor systems are, however, not expected to be able to trigger very strong bursts. Detailed chemical evolution models including SN I contributions and accounting for the finite stellar lifetimes (see Pap. I for details) give the abundance evolution of various elements from $^{12}C$ to $^{56}Fe$ for undisturbed galaxies of different spiral types.

In Fig. 1a we present the time evolution of the iron abundance $[Fe/H]$ and in Fig. 1b of the abundance ratio $[Mg/Fe]$ in spiral galaxies of various types. Starting from an initial value of $-12$, the ISM abundance $[Fe/H]$ climbs to values of $-1.2$ and $-1.35$ in $Sa$ and $Sb$ galaxies within 1 Gyr. Thereafter, the enrichment process slows down considerably and reaches $[Fe/H] = -0.2$ ($Sa$) and $[Fe/H] = -0.3$ ($Sb$) at ages of $12-15$ Gyr. In $Sc$ and $Sd$ galaxies, due to their longer SF timescale, the enrichment process is both weaker and slower, and $[Fe/H]$ only reaches values around $-0.7$ after $12 - 15$ Gyr. $[Mg/Fe]$ decreases from about 0.06 at 3 Gyr to values between 0.05 ($Sd$ and $Sc$) and 0.02 ($Sa$) after 12 Gyr.

**Fig. 1.** Time evolution of $[Fe/H]$ (**1a**) and of $[Mg/Fe]$ (**1b**) for spiral types $Sa$ through $Sc$ and for an $Sc - Sc$ merger occuring after 12 Gyr and triggering a burst of strength $b = 0.5$.

We have further studied the chemical evolution of a large variety of Sp – Sp merger events with their interaction – induced starbursts. For the sake of simplicity, we have confined our study to mergers of two spirals, identical in Hubble type and age. We find that during a burst, when the SF rate is increased over a time interval of the order of $10^8$ yr, the metallicity $Z$ increases rapidly (cf. Fig. 12 in Pap. I), reaching values $\gtrsim Z_\odot$ for burst strengths $b$ defined as the increase of stellar mass during the burst relative to the stellar mass at the onset of the burst – $b := \Delta S_{burst}/S_{preburst} \gtrsim 0.5$ in an $Sc - Sc$ merger taking place at 12 Gyr.

As an example we also present in Figs 1a and b the evolution of $[Fe/H]$ and $[Mg/Fe]$ in an $Sc$ galaxy pair, in which, after 12 Gyr of undisturbed evolution, a burst occurs as strong as to virtually consume all the available gas ($G/M \sim 40\%$). Such a strong burst makes $[Fe/H]$ increase from -0.73 to -0.28 on a timescale of $\sim 3 \cdot 10^8$ yr and decrease thereafter to -0.35 at 15 Gyr. With the onset of the burst, $[Mg/Fe]$ first goes down a little bit during some $10^7$ yr due to the fact that for stars in the mass range $15 - 40$ $M_\odot$ the iron yield is higher than the magnesium yield (we use SN nucleosynthesis from Woosley & Weaver (1986), see Pap. I). Then, the quickly returned SNII product Mg causes $[Mg/Fe]$ to rapidly increase from 0.045 to 0.065, whereafter, with the onset of SN I iron production, $[Mg/Fe]$ falls continuously and reaches a value of 0.035 at 15 Gyr (c.f. Fig. 1b).

The majority of the burst stars is expected to be borne with metallicity and abundance ratios close to those of the spiral progenitor ISM because first, the high metallicity phase during the burst is very short–lived and second, in a more realistic view than that of our simplified 1 – phase ISM models stars should form from a cold dense gas phase while the heavy elements produced by the first dying burst stars will be in hot low density SN bubbles. Only after a cooling time will these heavy elements again be available for astration. Cooling times may, however, be shorter in the highly non – equilibrium situations of strong bursts than in normal spiral galaxy SF mode. Thus, some self – enrichment effects cannot be excluded, in particular during long – lasting bursts. These may lead to an enhancement of SN II products like $^{16}O$ or $^{24}Mg$ with respect to elements originating from intermediate and low mass stars or type I SNe, as $^{12}C$ or $^{56}Fe$.

All this is, of course, particularly valid for any secondary population of GCs that is formed in the course of a starburst, induced by a merger of two spiral galaxies. Furthermore, it is expected to be valid when a massive gas – poor galaxy accretes a gas – rich companion and a starburst is triggered that only consumes the gas of the intruder. If two spirals of different Hubble types merge, the chemical evolution will be somehow intermediate between our respective equal type cases. We have shown in Pap. I that for mergers of $Sb$ through $Sd$ spirals occurring after 8 to 12 Gyr of undisturbed evolution, the metallicity of the burst stars and of any secondary GC population – if formed in the burst – should be $^1/_3 Z_\odot \lesssim Z \lesssim Z_\odot$ or $-0.8 \lesssim [Fe/H] \lesssim -0.2$. Observations of metallicity distributions of GC systems around suspected merger remnants as e.g. NGC 4472 and NGC 5128 indeed reveal a secondary peak within the range of our model predictions (cf. Harris et al. 1992, Zepf & Ashman 1993, Ostrov et al. 1993). GC systems around cD galaxies (e.g. NGC 1399) show a broad or multiply peaked metallicity distribution (Ostrov et al. 1993) as might be expected for these presumed multi - merger objects.

The resulting bimodal metallicity distribution of the final GC system will still testify to a merger origin of its parent galaxy when other suspicious features, like blue colours, tidal tails, etc., will long have disappeared.

## 4. Colour and line strength evolution for GCs of various metallicities

We model the photometric evolution of GCs for different metallicities using stellar evolutionary tracks and colour calibrations as described in Einsel et al. (1994). A Scalo – IMF with lower and upper mass limits of 0.15 and 60 $M_\odot$ is used together with a characteristic timescale $t_*$ for SF of $10^6$ yr for GCs. The colour and line strength evolution after the bulk of SF has occured does, however, not significantly depend on $t_*$ for $10^5 \leq t_* \leq 10^7$ yr. Stellar line strengths of Mgb and Fe5270 are calculated in our evolutionary synthesis models from the calibrations of Gorgas et al. (1993).

The colour evolution of the stellar population in GCs significantly depends on metallicity. This can be seen in Fig. 2a where, as one example only, we plot $V - I$ vs. time for different initial metallicities. Very crudely, one can say that going up in metallicity from $Z = 1 \cdot 10^{-4}$ to $1 \cdot 10^{-3}$, $1 \cdot 10^{-2}$, and

$4 \cdot 10^{-2}$, $V-I$ gets redder by $\sim 0.2$ mag per step for all stellar population ages from $\sim 1$ to 15 Gyr. The reason why the early ($\lesssim 10^8$ yr) $V-I$ colours for $Z = 2 \cdot Z_\odot$ fall close to those for $Z = 1/2 \cdot Z_\odot$ is that we had to use the same stellar evolutionary tracks for both metallicities (Einsel et al. 1994). For a 12 or 15 Gyr old population, the $V-I$ colours differ by as much as 0.6 mag between metallicities of 0.0001 and 0.04.

In $U-B$, the colour difference is only half as large at 0.1 Gyr, $\lesssim 0.1$ mag per step, as $Z$ climbs from $10^{-4}$ to $10^{-3}$, $10^{-2}$ and $4 \cdot 10^{-2}$, but it gets larger with age. At 12 to 15 Gyr, the $U-B$ colour difference is more than 1 mag between systems with $Z = 10^{-4}$ and those with $2 \cdot Z_\odot$. It should be mentioned that our GC models for half and twice solar metallicity fairly well bracket the $V-I$ evolution for solar metallicity as given by Bruzual & Charlot's (1993) models and depicted in Schweizer & Seitzer (1993).

Between evolutionary ages of $10^7$ and $10^{10}$ yr, e.g., a stellar system fades in U by 2.6 mag for $Z = 10^{-4}$ and by 4.2 mag for $Z = 10^{-2}$, in V by 1.5 mag ($Z = 10^{-4}$) and 2.7 mag ($Z = 10^{-2}$). As age increases from $10^7$ yr to 12 Gyr, the fading in K amounts to as much as 5.8 mag at $Z = 10^{-2}$ and to $\sim 2.5$ mag for $Z = 10^{-4}$.

In Figs 3a, b, we present the time evolution of Mgb and Fe5270 for GCs of 5 different metallicities. For both indices, our curves for $Z = 0.04$ and 0.01 nicely bracket the respective curves for solar metallicity presented by Schweizer & Seitzer using Bruzual & Charlot's models. At an evolutionary age of 13 Gyr, the difference between models with $Z = 10^{-4}$ and $Z = 2 \cdot Z_\odot$ amounts to as much as 3 Å in Mgb and 2 Å in Fe5270.

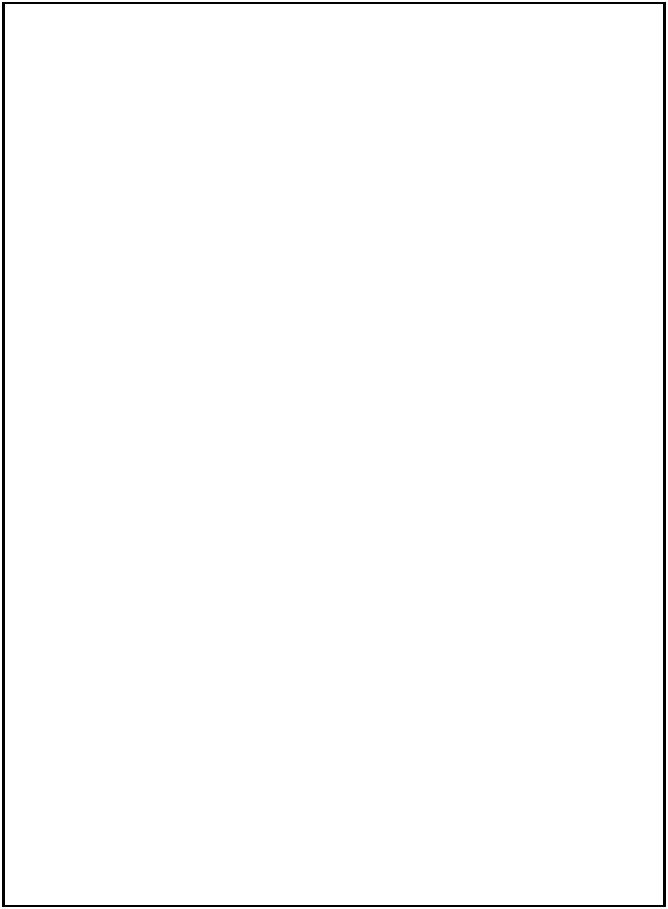

**Fig. 2.** Time evolution of $V-I$ (**2a**) and of $M_V$ (**2b**) for GCs of various metallicities.

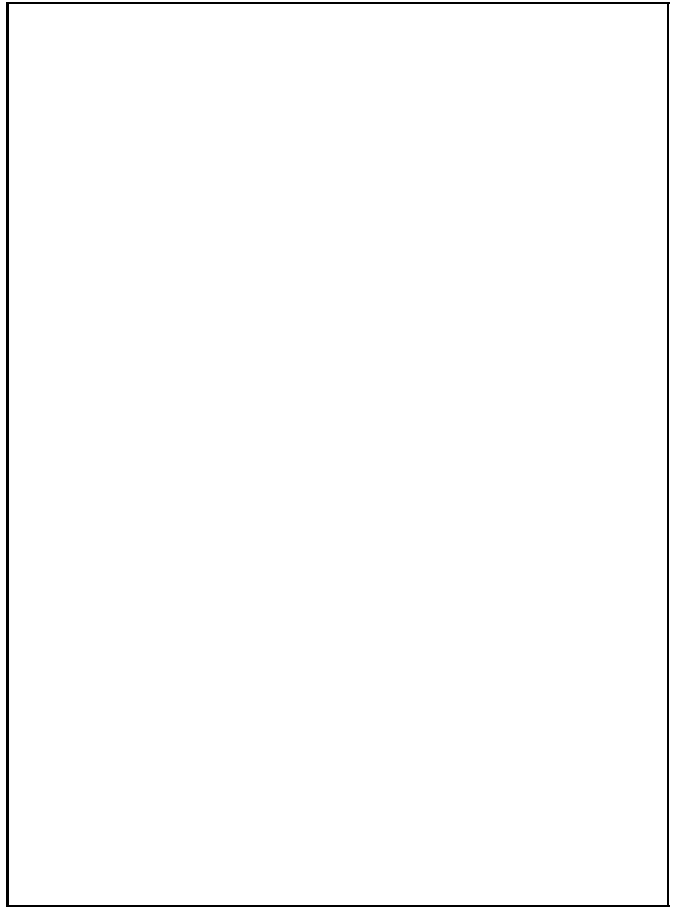

**Fig. 3.** Time evolution of Mgb (**3a**) and of Fe5270 (**3b**) for GCs of various metallicities.

In Fig. 2b, the luminosity evolution for different metallicities is presented on the example of $M_V$ for a stellar system which has $1.2 \cdot 10^9$ $M_\odot$ at 15 Gyr. If its metallicity is $10^{-3}$, our system fades by $\sim 4.5$ mag while going from 0.1 to 12 Gyr in age. In $U$ the fading is stronger and amounts to $\sim 5.7$ mag over the same age range, while in $I$ it's only 3.7 mag.

The amount of fading after 1 Gyr up to the present time is to good approximation no longer metallicity dependent, whereas in earlier phases, the fading strongly depends on $Z$.

## 5. GCs in NGC 7252

Whitmore et al. (1993) present $M_I$, $M_V$, $V-I$ and $R_{eff}$ for their sample of $\sim 40$ young GC candidates observed with HST around NGC 7252. They subdivide their sample into an outer one containing 36 objects at $\geq 3$ kpc from the center (we use $H_0 = 75$ km s$^{-1}$ Mpc$^{-1}$ throughout), and an inner one. The outer sample extends to $\sim 12$ kpc and is much more homogeneous in colour than the inner one which only contains outer sample clusters are typically around 10 pc, i.e. in the range of Galactic GC radii, while those of the inner sample are between 10 and 35 pc. Whitmore et al. take this as an indication that these inner sample objects might rather be open clusters or large associations than young GCs.

Spectra of the two brightest outer sample GCs, W3 and W30, have been taken with the Hale 5m telescope and are presented in Schweizer & Seitzer (1993). They cover the range from 4600 to 7000 Å and feature $H_\alpha$, $H_\beta$, Mgb and FeI+CaI$\lambda$5270 absorption features. Schweizer & Seitzer plan to take spectra for all of these young GCs in the near future, deeper HST observations of NGC 7252 are already scheduled and will reveal much fainter clusters.

Our code is presently extended to also model the spectroscopic evolution in a chemically consistent way using Kurucz's model atmosphere spectra for the various metallicities of our stellar evolutionary data base. So, in this first approach, we are only going to interpret the mean colours and luminosities of the outer (and inner) sample clusters and the two clusters W3 and W30 for which spectra are available, and wait for the future abservations to individually interpret the spectra of a larger number of GCs.

In a first step, let's assume that the young GCs in NGC 7252 have the metallicity predicted by our best fit spectral evolution model, $Z = 0.008$ (c.f. Pap. II). The dataset closest to this predicted metallicity has $Z = 10^{-2}$, i.e. half solar. The mean observed $V - I$ colour of the outer sample, $\langle V - I \rangle_{outer} = 0.83 \pm 0.25$ when compared to Fig. 2a then gives a mean age of $1.37^{+1.6}_{-0.8}$ Gyr for these outer sample GCs. This age of the secondary GC population agrees well with the burst age of $1.3^{+0.6}_{-0.1}$ Gyr which we obtained from our global spectrophotometric modelling of NGC 7252 and means that these blue young stellar systems with typical effective radii of GCs have really been formed during the strong merger induced burst that increased the total stellar population of this galaxy by 20 − 50 %. Dynamical considerations suggest that on average, these GCs should have been formed over the volume throughout which they are now observed, i.e. out to ∼ 12 kpc from the center. This confirms the global nature of the starburst and the enormous height of the SF efficiency over this volume, which had already been inferred from the radially constant colours of NGC 7252. Only high resolution HST imaging (Whitmore et al. 1993) has revealed a central blueing which may point to some late nuclear SF $\gtrsim$ 1 Gyr after the global burst.

The mean $V$ magnitude for these outer sample GCs as given by Whitmore et al. is $\langle V \rangle = 22.0 \pm 1.3$ With a distance modulus to NGC 7252 of $m - M = 34.00$ ($H_0 = 75$ km s$^{-1}$ Mpc$^{-1}$), this transforms into a mean absolute $\langle M_V \rangle = -12.0 \pm 1.3$. From their derived age of 1.3 Gyr to an assumed age of 13 (16) Gyr, comparable to our galactic GCs, they will fade by 2.8 (3.4) mag to $\langle M_V \rangle = -9.2(-8.6)$, respectively, and thus, almost reach the luminosity of $\omega$Cen. This amount of fading is almost independent of metallicity, as stated in Sect. 4. W3, by far the brightest among Whitmore et al. 's objects, will have $M_V \sim -13.3$ at an age of 13 Gyr and $M_V \sim -12.7$ at 16 Gyr. This is far beyond the luminosity range of galactic GCs, but probably not beyond the luminosity range of GCs in rich systems, as e.g. NGC 1399 (see Wagner et al. 1991). Super star clusters with luminosities comparable to or even higher than those of the GCs in NGC 7252 have also been detected in other merger remnants, like NGC 1140 (Hunter et al. 1994), NGC 1705 (O'Connell et al. 1994), and in the WR galaxy He 2 - 10 (Conti & Vacca 1994). The fact that all of the observed GCs will, if they survive for ∼ 13 − 16 Gyr, end up with luminosities $M_V \lesssim -7.3$, where $M_V \sim -7.3$ corresponds to maximum of the GC luminosity distribution function in the Milky Way or M31 (Harris & Racine 1979), seems to indicate that many more GCs than those observed by Whitmore et al. may have been formed in the merger of NGC 7252 in agreement with Ashman & Zepf's (1992) estimate.

For the two brightest young GC candidates with spectroscopy available, W3 and W30, the observed Mgb and Fe5270 can directly be compared to our Fig. 3. The observed value of $1.3 \pm 0.3$ Å for Fe5270 in W3 is reached by our curve for the $Z = 10^{-2}$ model at an age of $1.3^{+0.3}_{-0.6}$ Gyr in good agreement with the age estimate of $1.3^{+0.5}_{-0.1}$ Gyr based on its $V - I$ colour of $0.84 \pm 0.05$. The measured W(Mgb) = $3.5 \pm 0.5$ Å of W3, however, is only reached at an age of $6.9^{+2.4}_{-2.8}$ Gyr by our model clusters. At an assumed age of 1.3 Gyr our model for $Z = 10^{-2}$ has an Mgb value of only 1.5 Å and, at this age, a metallicity which is much higher than even $2 \cdot Z_\odot$ would be needed to account for the observed Mgb. Thus, if the Mgb value of W3 measured on a bright and locally variable galaxy background is not severely underestimated, it would mean that [Mg/Fe] is significantly enhanced. We have shown (cf. Fig. 12b in Pap. I) that in the course of strong starbursts as in NGC 7252 the global metallicity $Z$ – reflecting SNII products like $^{16}$O or $^{24}$Mg rather than $^{56}$Fe – can climb up to peak values $\sim 3 \cdot Z_\odot$ for a short time (cf. Fig. 12b in Pap. I). This leaves open the question of self-enrichment during the process of GC formation itself. Is the SN II product Mg produced as a result of self − enrichment during GC formation or as a result of a strong, global, long - lasting starburst (Fig.1b)?

Unfortunately, no Fe5270 could be given for the much bluer cluster W30 with $V - I = 0.53 \pm 0.05$ and W(Mgb) = $3.7 \pm 0.8$. So, we are left with considerable metallicity uncertainty and ensuing poor age restrictions. If W30 had the same metallicity as W3 – as suggested by its similar Mgb value – its age would come out to be $(2.7^{+1.3}_{-0.7}) \cdot 10^8$ yr. This would mean that W30 was formed long after the bulk global SF in NGC 7252. With a distance of only ∼ 5 kpc it is a little bit closer to the center of NGC 7252 than W3. If our scenario of a global starburst contracting to the center during its late phases (Pap. II) were correct, W30 could be formed during the late central afterburning. That the burst in NGC 7252 has a long and possibly strong SF tail extending until the present time within the innermost region seems indeed indicated by HST colour profiles and possibly also by ROSAT observations (c.f. Hibbard et al. 1993). Together with F. Schweizer we have IUE observations of NGC 7252 which should be able to answer this question. W30 could, however, also have a lower than half solar metallicity. Only for a metallicity as low as $Z \sim 5 \cdot 10^{-4}$ would its age come close to that of W3, the outer clusters and the burst age. In this case, [Mg/Fe] should be even more enhanced than for W3 in order to explain the observed Mgb. Clearly, observation of the Fe5270 line seems crucial to settle the age of this object.

## 6. Conclusions

In the strong starbursts that are triggered in the course of a merger of two massive gas-rich spirals the SFE over regions of

∼ 10 kpc size is high enough to allow for the formation of a second generation of GCs. Such a secondary GC population is readily identified by its enhanced metallicity even at stages where its colours and luminosities have already come close to those of typical old GCs. As shown on the example of NGC 7252, detailed chemical and spectrophotometric modelling of merging galaxies and merger remnants allows to rather precisely trace back the SF history and to predict the metallicity range of secondary GCs. We present evolutionary synthesis models for GCs in terms of broad band colours $UBVRIK$ and stellar metal indices Mgb and Fe5270, and show quantitatively the very strong metallicity dependence of broad band colour and line strength evolution.

For the predicted GC metallicity $Z = 0.008$ in NGC 7252 the mean $\langle V - I \rangle$ colour of some 40 of the young GCs detected by Whitmore et al. (1993) gives an age of $1.37^{+1.6}_{-0.8}$ Gyr in agreement with our global burst age estimate of $1.3^{+0.6}_{-0.1}$ Gyr.

For this age and metallicity the GCs will fade by 2.8 (3.4) mag in V and reach $M_V = -9.2(-8.6)$ at ages of 13 (16) Gyr.

For the two brightest GCs (W3 and W30) where spectroscopy (Schweizer & Seitzer 1993) is available, the Fe5270 index measured in one case (W3) strongly confirms our metallicity prediction and – thus removing the age-metallicity-degeneracy – allows a very accurate determination of this GC's age of $1.3^{+0.5}_{-0.1}$ Gyr. At an age of 13 – 16 Gyr, W3 will still have $M_V \sim -13$. This is brighter than $\omega$Cen, the brightest GC in the Milky Way, but probably not beyond the range of the most luminous GCs in rich GC systems. [Mg/Fe] seems significantly enhanced in W3. For W30 no Fe5270 is measured, so age determination is less accurate. $V - I$ alone yields an age of $(2.7^{+1.3}_{-0.7}) \cdot 10^8$ yr, younger than the burst and the other GCs. Perhaps W30 was produced in a long-lasting central starburst afterburning.

*The datafiles for the time evolution of $UBVRIK$ colours for GCs (i.e. single bursts of $10^5$ yr duration) of various metallicities can be obtained from ufritze at uni-sw.gwdg.de.*

*Acknowledgements.* We are grateful to F. Schweizer for valuable discussion, encouragement and the provision of his observational data prior to publication. We thank J. Truran for interesting discussions and the referee, Steve Zepf, for his very constructive criticism. U. F. – v. A. acknowledges financial support from the Verbundforschung Astronomie through BMFT grant WE-010 R 900-40 and from the SFB 328 (Galaxienentwicklung).